\documentclass[
 aps,
 ,twocolumn
]{revtex4}
\usepackage{graphicx}
\usepackage{amsmath}
\usepackage{xcolor}
\usepackage{latexsym}

\begin{document}

\title{Quantum Circuit Model for a Uniformly Accelerated Mirror}

\author{Daiqin Su$^{1}$}
\author{C. T. Marco Ho$^{1}$}
\author{Robert B. Mann$^{1,2,3}$}
\author{Timothy C. Ralph$^{1}$}
\affiliation{$^1$Centre for Quantum Computation and Communication Technology, School of Mathematics and Physics, University
of Queensland, Brisbane, Queensland 4072, Australia}
\affiliation{$^2$Perimeter Institute, 31 Caroline Steret North, Waterloo, Ontario N2L 2Y5, Canada}
\affiliation{$^3$ Department of Physics and Astronomy, University of Waterloo, Ontario, Canada, N2L 3G1}

\date{\today}


\begin{abstract}
We develop a quantum circuit model  describing unitary interactions between quantum fields
and a uniformly accelerated object, and apply it to  a semi-transparent mirror which uniformly accelerates in the Minkowski vacuum. 
The reflection coefficient $R_{\omega}$ of the mirror varies between 0 and 1, representing a generalization 
of the perfect mirror ($R_{\omega}=1$) discussed extensively in the literature. Our method is non-perturbative, not requiring  $R_{\omega} \sim 0$. We use the circuit model to calculate the 
radiation from an eternally accelerated mirror and obtain a finite particle flux along the past horizon provided an appropriate low frequency
regularization is introduced. More importantly, it is straightforward to see from our formalism that the radiation is squeezed. The 
squeezing is closely related to cutting the correlation across the horizon, which therefore may have important implications to the formation 
of a black hole firewall.
\end{abstract}


\maketitle

\vspace{10 mm}

\section{Introduction}

It has been well known since the 1970s that a moving mirror can radiate particles \cite{Moore70,Fulling76}.  A perfect moving mirror acts as a moving boundary and 
thus changes the states, especially the vacuum, of the quantum fields. For an appropriately chosen accelerated trajectory the radiation flux 
is thermal, and an analogy \cite{Davies77, Walker85, Carlitz87} can be drawn with Hawking radiation from a collapsing star \cite{Hawking75} that eventually forms a black hole. 
Since the thermal fluxes are correlated with the final vacuum fluctuations, 
some authors \cite{Wilczek93,Hotta15} have proposed that the emission of the large amounts of 
information left in the black hole need not be accompanied by the eventual emission of a large amount of energy, providing a new perspective to 
the solution of the black hole information paradox \cite{Hawking76}. 

The trajectory of a uniformly accelerated mirror is of  particular interest. 
When the mirror is uniformly accelerating, its trajectory is a hyperbola in spacetime, and both the energy flux and particle flux are zero \cite{Fulling76, Davies77, Birrell82, Grove86}. Particles and energy are only radiated when the acceleration of the 
mirror changes. In the case that the mirror eternally accelerates, the energy flux along the horizon is divergent \cite{Frolov79,Frolov80,Frolov99}. 
This divergence is evidently related to the ideal assumption that the mirror accelerates for infinitely long time. One way to get rid of the divergence
is to turn on and off the mirror so that effectively it interacts with the fields for a finite time \cite{Obadia01,Obadia03,Parentani03}. 

In this paper, we develop a quantum circuit model to describe unitary interactions between  quantum fields and a uniformly accelerated object (such as a mirror, cavity, squeezer {\it etc.}).
Our circuit model can be considered a further development
of the matrix formalism first proposed by Obadia and Parentani \cite{Obadia01} to describe a mirror following general trajectories. 
We concentrate on a uniformly accelerated object because the transformations between Minkowski modes, Rindler modes and 
Unruh modes are well known \cite{Unruh76,Takagi86,Crispino08} and can be represented by some simple quantum optical elements, like
two-mode squeezers, beamsplitters {\it etc.}  As an application of our circuit model, we 
revisit the uniformly accelerated mirror problem in $(1+1)$-dimensional Minkowski spacetime.
Unlike  the self-interaction model proposed by Obadia and Parentani \cite{Obadia01},
which requires a perturbative expansion and is valid only for low reflection coefficients, our circuit model is non-pertubative insofar as 
it is valid for any value of the reflection coefficient.

For the eternally accelerated mirror, the radiation flux in a localized wave packet mode is divergent. 
We can regularize this infrared divergence by introducing a low-frequency cutoff  for the mirror,
which means the mirror is transparent for the low-frequency field modes
(to some extent, this is physically equivalent to having the mirror interact with the field for a finite period of time). 
After  infrared regularization the particle number in a localized wave packet mode is finite. 
We further study the properties of 
the radiation flux and find that the radiation field is squeezed. This squeezing effect 
has gone unnoticed up to now, but in our circuit model it is a very
straightforward result. We show that the generation of squeezing is closely related to cutting the correlations across the horizon. This
mechanism of transferring correlations to squeezing may have important implications for black hole firewalls \cite{AMPS, Braunstein13}, as we shall subsequently discuss.


 
Our paper is organized as
follows. In Sec. \ref{RindlerUnruh}, we briefly review the relations between Rindler modes and Unruh modes.  
Motivated from these transformations, we introduce our circuit 
model in Sec. \ref{circuit} and calculate the radiation flux from an eternally accelerated mirror in Sec. \ref{radiation}. In Sec. \ref{squeezing},
we show that the radiation field from the accelerated mirror is squeezed and the squeezing is related to the 
correlations across the horizon. In Sec. \ref{firewall}, we propose that a Rindler firewall can be generated by a uniformly accelerated mirror
and we conjecture that a black hole firewall could be squeezed. We conclude in Sec. \ref{conclusion}. In this paper, we take the unit
$\hbar = c =1$. 



\section{Rindler modes and Unruh modes}\label{RindlerUnruh}

In this section we describe the relations between Rindler modes and Unruh modes, which act as the foundation of our quantum circuit model. We begin with a
 brief review of the three ways of quantizing a massless scalar field $\hat{\Phi}$ 
in $(1+1)$-dimensional Minkowski spacetime (for comprehensive reviews, see \cite{Takagi86, Crispino08}).

A massless scalar field $\hat{\Phi}$ satisfies the Klein-Gordon equation $\Box \hat{\Phi} = 0$, 
where the d' Alembertian $\Box \equiv  (\sqrt{-g})^{-1} \partial_{\mu} [\sqrt{-g}g^{\mu\nu} \partial_{\nu}]$ and 
$g_{\mu\nu}$ is the metric of the spacetime \cite{Birrell82}. 
In the inertial frame, Minkowski coordinates $(t, x)$ are used and the metric 
$g_{\mu\nu} = \eta_{\mu\nu} = \text{diag}\{-1, +1\}$. 
The scalar field $\hat{\Phi}$ can be quantized in the standard way, 
\begin{equation}\label{MinkowskiMode}
\hat{\Phi} = \int d k \big(\hat{a}_{1k} u_{1k}  + \hat{a}_{2k} u_{2k} +  \text{h.c.}  \big),
\end{equation}
where h.c. represents Hermitian conjugate, 
$u_{1k}~(u_{2k})$ are single-frequency left-moving (right-moving) mode functions
\begin{equation*}
u_{1k}(V) = (4 \pi k)^{-1/2} e^{-i k V}, \,\,\,\,\,\, u_{2k}(U) =(4 \pi k)^{-1/2} e^{-i k U}, 
\end{equation*}
with $V = t+x, U = t-x$.
$\hat{a}_{1k}(\hat{a}_{2k})$, $\hat{a}_{1k}^{\dag} (\hat{a}_{2k}^{\dag} )$
are the corresponding annihilation and creation operators satisfying the bosonic commutation relations
\begin{eqnarray*}
[\hat{a}_{mk}, \hat{a}_{nk'}^{\dag}] = \delta_{mn} \delta({k-k'}), \,\,\, [\hat{a}_{mk}, \hat{a}_{nk'}] = [\hat{a}_{mk}^{\dag}, \hat{a}_{nk'}^{\dag}] = 0, \nonumber
\end{eqnarray*}
with $m,n = 1,2$.  The Minkowski vacuum state $| 0_M \rangle$ is defined as $\hat{a}_{mk} | 0_M \rangle = 0$.
\begin{figure}[ht!]
\includegraphics[width=8.0cm]{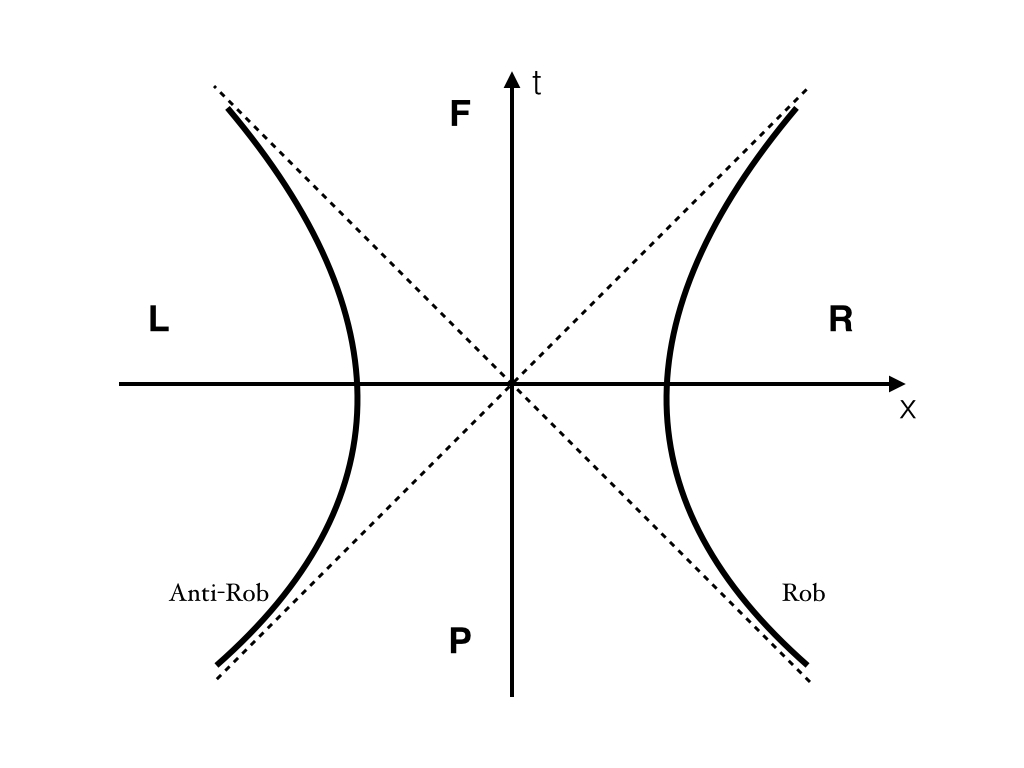}
\caption{\footnotesize Four wedges of $(1+1)$-dimensional Minkowski spacetime: $R, L, F$ and $P$. The right Rindler wedge ($R$) is causally disconnected 
to the left Rindler wedge ($L$). The Rindler coordinates $(\tau, \xi)$ only cover the $R$ wedge and $(\bar{\tau}, \bar{\xi})$ only
cover the $L$ wedge. } 
\label{RindlerFrame}
\end{figure}

As shown in Fig. \ref{RindlerFrame},  Minkowski spacetime can be divided into four wedges: $R, L, F$ and $P$.  
We introduce Rindler coordinates $(\tau, \xi)$ in the $R$ wedge and $(\bar{\tau}, \bar{\xi})$ in the $L$ wedge,
\begin{eqnarray}
t &=& a^{-1} e^{a \xi} \sinh(a \tau), \,\,\, x = a^{-1} e^{a \xi} \cosh(a \tau),  \nonumber \\
t &=& -a^{-1} e^{a \bar{\xi}} \sinh(a \bar{\tau}), \,\,\, x = - a^{-1} e^{a \bar{\xi}} \cosh(a \bar{\tau}),
\end{eqnarray}
where $\tau$ is the proper time of the uniformly accelerated observer with proper acceleration $a$
in the $R$ wedge. The metric is $g_{\mu\nu} = e^{2 a \xi} \text{diag}\{- 1, +1 \}$ in $R$
and is $g_{\mu\nu} = e^{2 a \bar{\xi}} \text{diag}\{- 1, +1 \}$ in $L$. It is obvious that the vector field
$\partial_{\tau}~(\text{or}~ \partial_{\bar{\tau}})$ is the timelike Killing vector field of the spacetime \cite{Birrell82}. 
In the Rindler frame, the scalar field $\hat{\Phi}$ can be quantized as \cite{Fulling73, Unruh76}
\begin{eqnarray}
\hat{\Phi} = \int d \omega (\hat{b}_{1\omega}^R g_{1\omega}^R   + \hat{b}_{1\omega}^L g_{1\omega}^L  
  + \hat{b}_{2\omega}^R g_{2\omega}^R  + \hat{b}_{2\omega}^L g_{2\omega}^L +  \text{h.c.} ) 
\end{eqnarray}
where the superscripts $``R"$ and $``L"$ represent modes and operators in the $R$ and $L$ wedge, respectively. The modes
$g_{m\omega}^R~(g_{m\omega}^L)$ only have support in the $R$ ($L$) wedge,
\begin{equation*}
g^R_{1\omega}(v) = (4 \pi \omega)^{-1/2} e^{-i \omega v}, \,\,\, g^R_{2\omega}(u) = (4 \pi \omega)^{-1/2} e^{-i \omega u},
\end{equation*}
where $v = \tau + \xi$, $u = \tau - \xi$,  and by replacing $v, u$ by $\bar{v} = - \bar{\tau} - \bar{\xi}$ and $\bar{u} = - \bar{\tau} + \bar{\xi}$
we obtain modes in the $L$ wedge. Note that we have used the prescription that $\partial_{\bar{\tau}}$ is past-directed.
The commutation relations of the operators are
\begin{eqnarray*}
[\hat{b}_{m\omega}^R, \hat{b}_{n\omega'}^{R\dag} ] = \delta_{mn} \delta({\omega-\omega'}), \,\,\,\,
[\hat{b}_{m\omega}^L, \hat{b}_{n\omega'}^{L\dag} ] = \delta_{mn} \delta({\omega-\omega'}), 
\end{eqnarray*}
with all others vanishing. The Rindler vacuum state $| 0_R \rangle$ is defined as 
$\hat{b}_{m\omega}^R | 0_R \rangle = \hat{b}_{m\omega}^L | 0_R \rangle =0$.

It proves  useful to introduce Unruh modes (instead of Minkowski modes) that cover the whole Minkowski spacetime for 
two reasons: 1) the Unruh and Minkowski modes share the same vacuum;  2) the transformation between Rindler modes
and Unruh modes is a two-mode squeezing transformation. The Unruh modes are defined as 
\begin{eqnarray}\label{UnruhRindler}
\hat{c}_{m\omega} &=& \text{cosh}(r_{\omega}) \hat{b}_{m\omega}^R - \text{sinh}(r_{\omega}) \hat{b}_{m\omega}^{L\dag}, \nonumber \\
\hat{d}_{m\omega} &=& \text{cosh}(r_{\omega}) \hat{b}_{m\omega}^L - \text{sinh}(r_{\omega}) \hat{b}_{m\omega}^{R\dag},
\end{eqnarray}
where $r_{\omega}$ satisfies $\text{tanh}(r_{\omega}) = e^{-\pi \omega/a}$. 
It is easy to find the inverse transformation,
\begin{eqnarray}\label{UnruhRindler:inverse}
\hat{b}_{m\omega}^R &=& \text{cosh}(r_{\omega}) \hat{c}_{m\omega} + \text{sinh}(r_{\omega}) \hat{d}_{m\omega}^{\dag}, \nonumber\\
\hat{b}_{m\omega}^L &=& \text{cosh}(r_{\omega}) \hat{d}_{m\omega} + \text{sinh}(r_{\omega}) \hat{c}_{m\omega}^{\dag}. 
\end{eqnarray}
We can see that the Rindler modes $(\hat{b}_{m\omega}^R, \hat{b}_{m\omega}^L)$ and Unruh modes $(\hat{c}_{m\omega}, \hat{d}_{m\omega} )$ 
are related by a two-mode squeezing operator with a frequency dependent squeezing parameter $r_{\omega}$. In terms of Unruh modes,
the scalar field $\hat{\Phi}$ can be expressed as
\begin{eqnarray}\label{UnruhMode}
\hat{\Phi} &=& \int d \omega (\hat{c}_{1\omega} G_{1\omega} 
  + \hat{d}_{1\omega} \bar{G}_{1\omega} 
   \nonumber \\
&&\qquad\qquad + \hat{c}_{2\omega} G_{2\omega}  + \hat{d}_{2\omega} \bar{G}_{2\omega} + \text{h.c.} 
  )
\end{eqnarray}
where
\begin{eqnarray}\label{Unruhmodes}
G_{1\omega}(V) &=& F(\omega, a) (aV)^{-i \omega/a},  \nonumber\\
\bar{G}_{1\omega}(V) &=& F(\omega, a) (-aV)^{i \omega/a}, \nonumber\\
G_{2\omega}(U) &=& F(\omega, a) (-aU)^{i \omega/a}, \nonumber\\
\bar{G}_{2\omega}(U) &=& F(\omega, a) (aU)^{-i \omega/a},
\end{eqnarray}
with $F(\omega, a) \equiv \frac{e^{\pi \omega/2a}}{\sqrt{4\pi \omega} \sqrt{2\sinh(\pi \omega/a)}}$.
$G_{1\omega}(V)$ and $\bar{G}_{2\omega}(U)$ are analytic in the lower-half complex plane while 
$\bar{G}_{1\omega}(V)$ and $G_{2\omega}(U)$ are analytic in the upper-half complex plane. 
The Unruh modes annihilate the Minkowski vacuum state
\begin{eqnarray*}
\hat{c}_{m\omega} | 0_M \rangle = \hat{d}_{m\omega} | 0_M \rangle = 0
\end{eqnarray*}
as noted above.


\section{Circuit model}\label{circuit}

\subsection{General formalism}

How are the states of a quantum field affected by an object (such as a  beamsplitter) that is 
uniformly accelerated in the $R$ wedge?  This is the question of central interest in this paper.
A straightforward way to study this problem is to work in the accelerated 
frame in which the object is static. It is obvious that the object only interacts with Rindler modes in the $R$ wedge and 
the Rindler modes in the $L$ wedge remain unaffected.  The interaction between the object and the Rindler
modes is unitary and it transforms the Rindler modes as
\begin{equation}\label{generalcoupling}
\hat{b}_{mk}^{\prime R} = \int d \omega \bigg( \alpha^{m1}_{k \omega} \hat{b}_{1\omega}^R 
+ \beta^{m1}_{k \omega} \hat{b}_{1\omega}^{R \dagger} 
+ \alpha^{m2}_{k \omega} \hat{b}_{2 \omega}^R 
+ \beta^{m2}_{k \omega} \hat{b}_{2\omega}^{R \dagger} \bigg). \\
\end{equation} 
This is the most general interaction which not only couples the left-moving and right-moving Rindler modes but also 
Rindler modes with different frequencies. Together with Eqs. (\ref{UnruhRindler}) and (\ref{UnruhRindler:inverse}), we 
can construct a quantum circuit model (or  input-output formalism) for the uniformly accelerated object. We start from the inertial frame in which
Unruh modes are used instead of Minkowski modes. This makes the model simpler although we still have to transform
the Minkowski modes to the Unruh modes and vice versa. 

First, based on Eq. (\ref{UnruhRindler:inverse}), the Unruh modes 
pass through a collection of two-mode squeezers each of which couples a pair of Unruh modes $(\hat{c}_{m\omega}, \hat{d}_{m\omega} )$
with frequency dependent squeezing parameter $r_{\omega}$. Second, the output right Rindler modes $\hat{b}_{m\omega}^R $
interact with the object and are transformed to $\hat{b}_{mk}^{\prime R}$ according to Eq. (\ref{generalcoupling}) while the left 
Rindler modes $\hat{b}_{m\omega}^L$ remain unchanged. Finally, based on Eq. (\ref{UnruhRindler}),  the Rindler modes 
pass through a collection of two-mode antisqueezers and are transformed to output Unruh modes $(\hat{c}^{\prime}_{m\omega}, \hat{d}^{\prime}_{m\omega} )$. 
If we use an inertial detector to detect the radiation field from the accelerated object, we have to transform the Unruh modes
$(\hat{c}^{\prime}_{m\omega}, \hat{d}^{\prime}_{m\omega} )$ to Minkowski modes to model the coupling with the detector.

\begin{figure}[ht!]
\includegraphics[width=8.0cm]{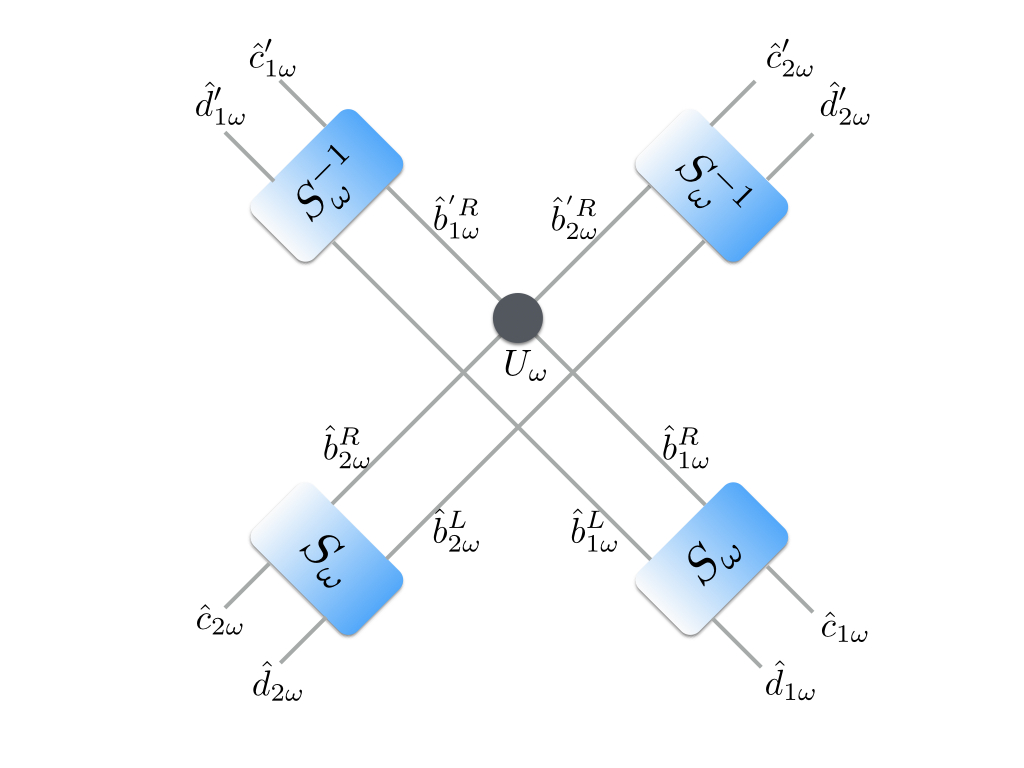}
\caption{\footnotesize (color online). Unruh modes pass through the squeezers and then become Rindler modes. The Rindler modes in the right Rindler 
wedge interact with the object ($U_{\omega}$) and then combine with the Rindler modes from the left Rindler wedge in the antisqueezers, 
going back to Unruh modes again. } 
\label{circuit:Fig}
\end{figure}

In the special case that the interaction does not couple Rindler modes with different frequencies, the input-output formalism is substantially 
simplified.  The coefficients $\alpha^{mn}_{k \omega}$ and $\beta^{mn}_{k \omega}$ are now
proportional to $\delta(k - \omega)$ so Eq. (\ref{generalcoupling}) becomes 
\begin{equation}\label{nomixingfrequency}
\hat{b}_{m\omega}^{\prime R} = \alpha^{m1}_{\omega \omega} \hat{b}_{1\omega}^R 
+ \beta^{m1}_{\omega \omega} \hat{b}_{1\omega}^{R \dagger} 
+ \alpha^{m2}_{\omega \omega} \hat{b}_{2 \omega}^R 
+ \beta^{m2}_{\omega \omega} \hat{b}_{2\omega}^{R \dagger}.
\end{equation} 
Since modes with different frequencies are independent, we can propose a quantum circuit model for each single frequency. The 
quantum circuit is shown in Fig.  \ref{circuit:Fig}.
A pair of left-moving Unruh modes $(\hat{c}_{1\omega}, \hat{d}_{1\omega})$ and a pair of right-moving Unruh modes $(\hat{c}_{2\omega}, \hat{d}_{2\omega})$
pass through the two-mode squeezers $S_{\omega}$, from which emerge left-moving Rindler modes $(\hat{b}_{1\omega}^R, \hat{b}_{1\omega}^L)$ and right-moving
Rindler modes $(\hat{b}_{2\omega}^R, \hat{b}_{2\omega}^L)$, respectively. $\hat{b}_{1\omega}^R$ and $\hat{b}_{2\omega}^R$
interact with each other when passing through the object (symbolized by the black dot in Fig. \ref{circuit:Fig}) and emerge as $\hat{b}_{1\omega}^{\prime R}$ and $\hat{b}_{2\omega}^{\prime R}$, 
which can be described by a unitary transformation $U_{\omega}$ according to Eq. (\ref{nomixingfrequency}). After that, the 
Rindler modes are combined by two-mode antisqueezers $S^{-1}_{\omega}$, ending up with Unruh modes again. 

For computational purposes, we introduce operator vectors $\hat{\bf c}_{\omega}$, $\hat{\bf d}_{\omega}$, $\hat{\bf b}_{\omega}^R$ and
$\hat{\bf b}_{\omega}^L$, which are defined as
\begin{eqnarray*}
\hat{\bf c}_{\omega} = {\hat{c}_{\omega} \choose \hat{c}_{\omega}^{\dag}}, \,\,\,\, \hat{\bf d}_{\omega} = {\hat{d}_{\omega} \choose \hat{d}_{\omega}^{\dag}}, \,\,\,\,
\hat{\bf b}_{\omega}^R = {\hat{b}_{\omega}^R \choose \hat{b}_{\omega}^{R\dag}},\,\,\,\, \hat{\bf b}_{\omega}^L = {\hat{b}_{\omega}^L \choose \hat{b}_{\omega}^{L\dag}}.
\end{eqnarray*}
Then Eqs. (\ref{UnruhRindler}) and (\ref{UnruhRindler:inverse}) can be rewritten as
\begin{eqnarray}
{\hat{\bf c}_{m\omega} \choose \hat{\bf d}_{m\omega}} = S^{-1}_{\omega} {\hat{\bf b}_{m\omega}^R \choose \hat{\bf b}_{m\omega}^L}, \,\,\,\,\,\,
{\hat{\bf b}_{m\omega}^R \choose \hat{\bf b}_{m\omega}^L} =S_{\omega}{\hat{\bf c}_{m\omega} \choose \hat{\bf d}_{m\omega}},
\end{eqnarray}
with
\begin{eqnarray}
S_{\omega} \equiv \left({I \text{cosh}(r_{\omega}) \atop \sigma_x \text{sinh}(r_{\omega})}{ \sigma_x \text{sinh}(r_{\omega}) 
\atop I \text{cosh}(r_{\omega})}\right)
\end{eqnarray}
where $I = \left({1 \atop 0}{0 \atop 1}  \right)$ is the identity matrix and $\sigma_x = \left({0 \atop 1}{1 \atop 0} \right)$ is one of the Pauli matrices.
The transformation between the input Unruh modes $(\hat{\bf c}_{1\omega}, \hat{\bf d}_{1\omega}, \hat{\bf c}_{2\omega}, \hat{\bf d}_{2\omega})^T$ and 
the output Unruh modes $(\hat{\bf c}^{\prime}_{1\omega} \hat{\bf d}^{\prime}_{1\omega}, \hat{\bf c}^{\prime}_{2\omega}, \hat{\bf d}^{\prime}_{2\omega})^T$ can be 
represented as
\begin{eqnarray}\label{input-output}
\begin{pmatrix}
\hat{\bf c}^{\prime}_{1\omega} \\ \hat{\bf d}^{\prime}_{1\omega} \\ \hat{\bf c}^{\prime}_{2\omega}  \\ \hat{\bf d}^{\prime}_{2\omega}
\end{pmatrix}
=\mathcal{S}^{-1}_{\omega} \mathcal{U}_{\omega} \mathcal{S}_{\omega}
\begin{pmatrix}
\hat{\bf c}_{1\omega} \\ \hat{\bf d}_{1\omega} \\ \hat{\bf c}_{2\omega} \\ \hat{\bf d}_{2\omega}
\end{pmatrix}.
\end{eqnarray}
$\mathcal{S}_{\omega}$ characterizes the transformation from Unruh modes to Rindler modes
\begin{eqnarray}
\mathcal{S}_{\omega} = \left({S_{\omega} \atop 0}{ 0 
\atop S_{\omega}}\right)
\end{eqnarray}
and $\mathcal{U}_{\omega}$ characterizes the action of the object 
\begin{eqnarray}
\mathcal{U}_{\omega} = 
\begin{pmatrix}
U^{11}_{\omega} & 0 & U^{12}_{\omega}  & 0 \\ 
0 & I  & 0 & 0 \\ 
U^{21}_{\omega} & 0 & U^{22}_{\omega}  & 0\\ 
0 & 0 & 0 & I 
\end{pmatrix}
\end{eqnarray}
where 
\begin{eqnarray}
U^{mn}_{\omega} = \left({\alpha^{mn}_{\omega \omega} \atop \beta^{mn*}_{\omega \omega}}{ \beta^{mn}_{\omega \omega}
\atop \alpha^{mn*}_{\omega \omega}}\right).
\end{eqnarray}
We emphasize that the general formalism developed here is valid for a wide class of quantum optical devices (objects), such as beamsplitters, 
single-mode squeezers, two-mode squeezers, cavities, and even for devices with time-dependent parameters, for example, beamsplitters
with time-dependent transmission coefficients. In this paper, we mainly apply the formalism to the simplest case, a beamsplitter.

\subsection{Circuit model for a uniformly accelerated mirror}


The perfect moving mirror problem has been extensively studied for several decades. 
A perfect moving mirror provides a clear boundary for a quantum field, which vanishes along the mirror's trajectory. 
The standard method for calculating the radiation from a perfect moving 
mirror is to find the Bogoliubov transformation between the input and output modes by taking into account the Dirichlet boundary 
condition. 

However a realistic mirror is not perfect but usually partially transparent, for which the Dirichlet 
boundary condition is not satisfied. In this paper, we are interested in a uniformly accelerated imperfect mirror whose motion looks nontrivial 
for an inertial observer. Rather than use the standard method (which is still valid if appropriate boundary conditions are considered), 
we shall employ the circuit model developed in the previous section, leading to a much simpler way to attack this problem. 




The idea is to work in the accelerated frame, in which the mirror is static and can be considered as a beamsplitter.
Without loss of generality, we assume that the mirror uniformly accelerates in the $R$ wedge. 
The beamsplitter transforms the right Rindler modes as
\begin{eqnarray}\label{beamsplitter}
\hat{b}^{\prime R}_{1\omega} &=&\text{cos}\,\theta_{\omega} \hat{b}^R_{1\omega}- i e^{i \phi_{\omega}} \text{sin}\,\theta_{\omega} \hat{b}^R_{2\omega}, \nonumber\\
\hat{b}^{\prime R}_{2\omega} &=&\text{cos} \,\theta_{\omega}  \hat{b}^R_{2\omega}- i e^{-i \phi_{\omega}} \text{sin}\,\theta_{\omega} \hat{b}^R_{1\omega},
\end{eqnarray}
where $\theta_{\omega}$ an $\phi_{\omega}$ are frequency dependent. The relative phase shift $i e^{\pm i \phi_{\omega}} $ ensures that the 
transformation is unitary. The intensity reflection and transmission coefficients of the beamsplitter are
\begin{eqnarray*}
R_{\omega} = \text{sin}^2\,\theta_{\omega}, \,\,\,\,\,\,\,\, T_{\omega} =  \text{cos}^2\,\theta_{\omega}. 
\end{eqnarray*}
By comparing Eqs. (\ref{beamsplitter}) and (\ref{nomixingfrequency}) we have 
\begin{eqnarray*}
\alpha^{11}_{\omega \omega} &=& \alpha^{22}_{\omega \omega} = \cos \theta_{\omega},  \nonumber \\
\alpha^{12}_{\omega \omega} &=& - \alpha^{21*}_{\omega \omega} = -i e^{i \phi_{\omega}} \sin \theta_{\omega},
\end{eqnarray*}
and all $\beta^{mn}_{\omega \omega}$ are zero.
We can therefore express the action of the beamsplitter as
\begin{eqnarray}\label{mirrortransformation}
\mathcal{U}_{\omega} = 
\begin{pmatrix}
I \text{cos}\,\theta_{\omega}& 0 & Z \text{sin}\,\theta_{\omega}  & 0 \\ 
0 & I  & 0 & 0 \\ 
-Z^* \text{sin}\,\theta_{\omega} & 0 & I \text{cos}\,\theta_{\omega}  & 0\\ 
0 & 0 & 0 & I 
\end{pmatrix},
\end{eqnarray}
where  $I$ is the $2\times 2$ identity matrix and
\begin{equation}
Z = \left({-i e^{i \phi_{\omega}} \atop 0}{0 \atop i e^{-i \phi_{\omega}}} \right).
\end{equation}

The explicit expressions for the transformation Eq. (\ref{input-output}) can be calculated straightforwardly and are summarized in  
Appendix \ref{appendixA}. With these transformations, it is easy to calculate the expectation value of the particle number of the output
$\hat{c}^{\prime}_{1\omega}$,
\begin{eqnarray}\label{ParticleNumber:1}
&&\langle 0_M | \hat{c}_{1\omega}^{\prime \dag} \hat{c}_{1\omega^{\prime}}^{\prime} | 0_M \rangle 
\nonumber \\
&=& 2(1-\text{cos} \,\theta_{\omega})\text{cosh}^2(r_{\omega}) \text{sinh}^2(r_{\omega}) \delta(\omega-\omega^{\prime}) \nonumber \\
&=& 2(1-\text{cos} \,\theta_{\omega})\frac{e^{2\pi \omega/a}}{(e^{2\pi \omega/a}-1)^2} \delta(\omega-\omega^{\prime}) \nonumber \\
&\equiv& n(\omega) \delta(\omega-\omega^{\prime}).
\end{eqnarray}
The corresponding expectation values for the other three outputs is the same as Eq. (\ref{ParticleNumber:1}). 
Hence the number of Unruh particles in every output is generally not zero. The particle-number distribution is 
\begin{equation}\label{ParticleNumber:2}
n(\omega) = 2 (1-\text{cos}\,\theta_{\omega})\frac{e^{2\pi \omega/a}}{(e^{2\pi \omega/a}-1)^2},
\end{equation}
depending on the transmission coefficient of the uniformly accelerated mirror. Note that 
$n(\omega) = 0$ only when $\theta_{\omega} = 0$; in other words when the mirror is completely transparent to the field mode with frequency $\omega$.  
We also note that the distribution of the output Unruh particles is not thermal.


\section{Radiation from an eternally accelerated mirror}\label{radiation}

\begin{figure}[ht!]
\includegraphics[width=8.0cm]{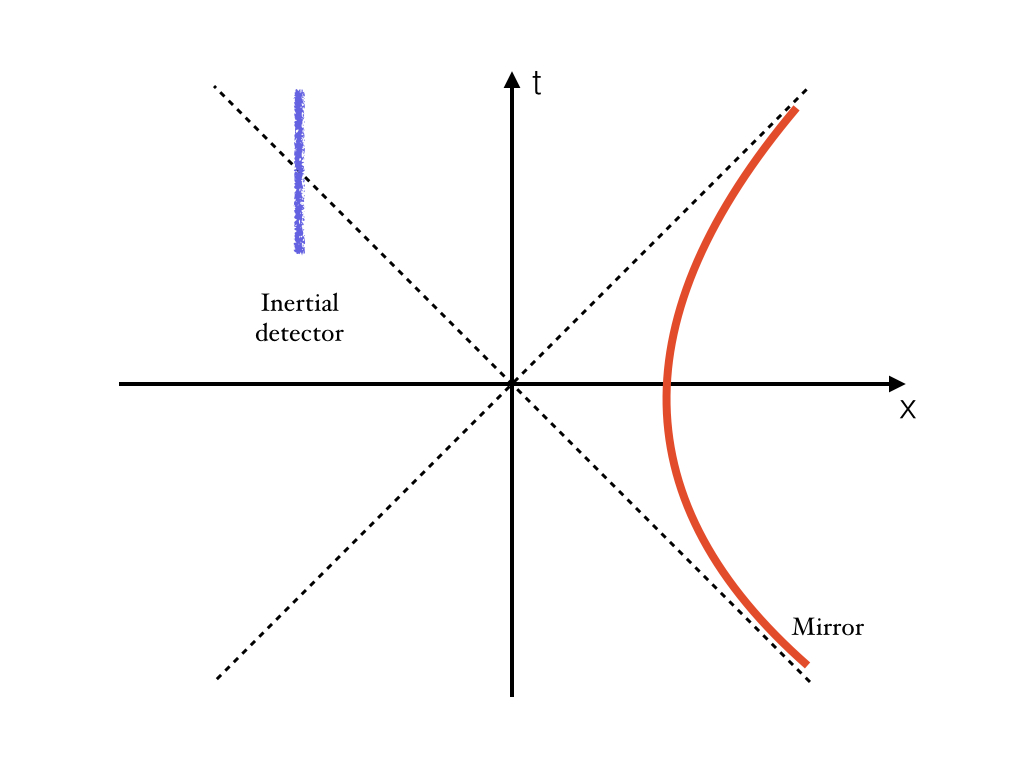}
\caption{\footnotesize (color online). A uniformly accelerated mirror on the right Rindler wedge. An inertial detector is placed at an appropriate position 
to detect left-moving particles coming from the uniformly accelerated mirror. } 
\label{Fig:Mirror}
\end{figure}
As an application of the quantum circuit model, we calculate the radiation flux from an eternally accelerated mirror. As shown in Fig. \ref{Fig:Mirror},
an inertial detector is placed at an appropriate position to detect the left-moving particles radiated by the accelerated mirror. 
In the previous section, we have shown that the accelerated mirror radiates Unruh particles. However, the inertial detector responds only to Minkowski particles. 
In order to calculate the response of the inertial detector we need to find the transformation between Unruh modes and Minkowski modes. 
This can be done by comparing Eqs. (\ref{MinkowskiMode}) and (\ref{UnruhMode}), and then using the Klein-Gordon inner product \cite{Birrell82}, 
\begin{eqnarray}\label{MinkowskiUnruh}
\hat{a}_k &=& \int d \omega \bigg(  \langle u_k, G_{\omega}  \rangle\hat{c}^{\prime}_{\omega} + \langle u_k, \bar{G}_{\omega}  \rangle\hat{d}^{\prime}_{\omega} \bigg) \nonumber \\
&\equiv& \int d \omega ( A_{k \omega} \hat{c}^{\prime}_{\omega} + B_{k \omega} \hat{d}^{\prime}_{\omega} ),
\end{eqnarray}
where $A_{k\omega} =  \langle u_k, G_{\omega}  \rangle$ and $B_{k\omega} = \langle u_k, \bar{G}_{\omega}  \rangle$ are the 
Bogoliubov transformation coefficients. Since we only consider left-moving modes here, without
introducing any confusion,  we have omitted the subscript $``1"$. Using the relation between Unruh modes and Rindler modes
Eq. (\ref{UnruhRindler}) and the relation between Rindler modes and Minkowski modes \cite{Crispino08}, We can find the 
transformation between Unruh modes and Minkowski modes. A more straightforward way is to directly calculate the Klein-Gordon inner product using
the explicit expressions of Unruh modes Eq. (\ref{Unruhmodes}). The result is 
\begin{eqnarray}
A_{k\omega} = B^*_{k\omega} =  \frac{i \sqrt{2 \sinh(\pi \omega/a)}}{2\pi \sqrt{\omega k}} \Gamma(1-i\omega/a) \bigg(\frac{k}{a}\bigg)^{i\omega/a}, \nonumber \\
\end{eqnarray}
where $\Gamma(z)$ is the Gamma function. 
In realistic quantum optics experiments a detector normally detects localized wave packet modes. 
In order to take this into account we consider Gaussian wave packet modes defined as
\begin{equation}\label{GaussianMode}
\hat{a}(f) = \int_0^{\infty} d k f(k; k_0, \sigma, V_0) \hat{a}_k, 
\end{equation}
where
\begin{equation}
f(k; k_0, \sigma, V_0) = \bigg(\frac{1}{2\pi \sigma^2}  \bigg)^{1/4} \exp\bigg\{-\frac{(k-k_0)^2}{4\sigma^2} - i k V_0\bigg\}
\end{equation}
with  $k_0$, $\sigma$ and $V_0$ the central frequency, bandwidth and central position, respectively. In the narrow bandwidth limit ($k_0 \gg \sigma$), 
the integration over $k$ can be approximately calculated to a very good accuracy. 

When $k_0 \gg \sigma$, the Gaussian wave
packet $f(k; k_0, \sigma, V_0)$ is significantly nonzero only for positive $k$, so the range of integration of $k$ can be extended to $(-\infty, \infty)$
without introducing large errors. Secondly, since $f(k; k_0, \sigma, V_0)$ is well localized around $k_0$, those values of $A_{k\omega}$ and $B_{k\omega}$ only
near $k_0$ are relevant. Writing \cite{Downes13}
\begin{equation}
\frac{1}{\sqrt{k}}\bigg( \frac{k}{a}\bigg)^{i \omega/a} \approx \frac{1}{\sqrt{k_0}}e^{i\frac{\omega}{k_0}\frac{k}{a}}e^{i\frac{\omega}{a}[\ln(\frac{k_0}{a})-1]}
\end{equation}
and then  expanding $A_{k\omega}$ and $B_{k\omega}$ around $k_0$ yields
\begin{widetext}
\begin{eqnarray}\label{A}
A_{\omega} \equiv \int_0^{\infty} d k f(k) A_{k \omega} 
&\approx& i \sqrt{\frac{\sigma}{\pi \omega k_0}} \bigg(\frac{1}{2\pi}  \bigg)^{1/4} \sqrt{2\sinh(\pi \omega/a)} \Gamma(1-i\omega/a)
e^{i\frac{\omega}{a}\ln(\frac{k_0}{a})} e^{-i k_0 V_0}\exp\bigg\{-\frac{\sigma^2(\omega/a-k_0 V_0)^2}{k_0^2}\bigg\},  \nonumber \\
\end{eqnarray}
\begin{eqnarray}\label{B}
B_{\omega} \equiv \int_0^{\infty} d k f(k) B_{k \omega}
&\approx& -i \sqrt{\frac{\sigma}{\pi \omega k_0}} \bigg(\frac{1}{2\pi}  \bigg)^{1/4} \sqrt{2\sinh(\pi \omega/a)} \Gamma(1+i\omega/a)
e^{-i\frac{\omega}{a}\ln(\frac{k_0}{a})} e^{-i k_0 V_0}\exp\bigg\{-\frac{\sigma^2(\omega/a+k_0 V_0)^2}{k_0^2}\bigg\} \nonumber \\
\end{eqnarray}
\end{widetext}
up to first order in $k-k_0$.

Using Eq. (\ref{ParticleNumber:1}) and
\begin{equation}
|\Gamma(1-i\omega/a)|^2 = |\Gamma(1+i\omega/a)|^2  = \frac{\pi \omega/a}{\sinh(\pi \omega/a)}
\end{equation} 
the expectation value  $N(f) = \langle 0_M | \hat{a}^{\dagger}(f)\hat{a}(f)|0_M \rangle$ 
 of  the Gaussian mode particle number  is
\begin{widetext}
\begin{eqnarray}\label{ParticleNumber}
N(f) &=& \int d \omega \int d \omega^{\prime}
\langle 0_M |(A^*_{\omega} \hat{c}^{\prime \dagger}_{\omega} +B^*_{\omega}\hat{d}^{\prime \dagger}_{\omega})
(A_{\omega^{\prime}} \hat{c}^{\prime}_{\omega^{\prime}} +B_{\omega'}\hat{d}^{\prime}_{\omega^{\prime}}) |0_M \rangle \nonumber \\
\nonumber \\
&=& 2 \int d \omega (|A_{\omega}|^2+|B_{\omega}|^2)(1-\cos \theta_{\omega})\frac{e^{2\pi \omega/a}}{(e^{2\pi \omega/a}-1)^2}, \nonumber \\
\nonumber \\
&=& \sqrt{\frac{8}{\pi}}\frac{\sigma}{k_0} \int_0^{\infty} d \Omega 
\bigg\{\exp\bigg[-\frac{2\sigma^2(\Omega-k_0 V_0)^2}{k_0^2}\bigg]+\exp\bigg[-\frac{2\sigma^2(\Omega+k_0 V_0)^2}{k_0^2}\bigg] \bigg\}
(1-\cos \theta_{\Omega})\frac{e^{2\pi \Omega}}{(e^{2\pi \Omega}-1)^2}
\end{eqnarray}
\end{widetext}
where $\Omega = \omega/a$ is the dimensionless Rindler frequency.  

Two special cases are of particular interest. 
Consider first that the mirror is completely transparent for all modes, that is 
$\cos^2 \, \theta_{\omega} = 1$. From Eq. (\ref{ParticleNumber}),  the particle number
vanishes, $N(f)=0$. This is not surprising because a completely transparent mirror does nothing to the Minkowski vacuum.
The second case is that the mirror is perfect for all modes, that is, $\cos^2 \, \theta_{\omega} = 0$. 
When $\Omega \rightarrow 0$, $(e^{2\pi \Omega}-1)^{-2} \sim \Omega^{-2} $ and all other factors in the integrand of 
Eq. (\ref{ParticleNumber}) are finite. Therefore, the particle number $N(f)$ is divergent.

This infrared divergence occurs because we naively assume that the mirror accelerates for an infinitely long time, which seems physically unreasonable. In the framework of the self-interaction model, the mirror is switched on and off so that one obtains finite particle flux \cite{Obadia01}. In our circuit model, we could also switch on and off the
mirror.  However instead   we shall use a simpler method of regularization. The idea is to directly introduce a low frequency
cutoff for the mirror, that is, the mirror is completely transparent for low-frequency field modes. The mechanism for a 
physical mirror to reflect electromagnetic waves is that the atoms consisting of the mirror absorb electromagnetic waves
and then reemit them again. If the wavelength of the electromagnetic wave is very long, the response time of the mirror is very long. Hence if the mirror accelerates for a finite time, it cannot respond to Rindler modes with characteristic 
oscillation period longer than the accelerating time. 

In this sense, introducing a low-frequency cutoff is equivalent to 
switching on and off the mirror. In higher dimensional spacetime, e.g., $(1+3)$-dimensional spacetime, there is another
reason justifying a low-frequency cutoff.  A physical mirror with finite size cannot reflect
field modes whose wavelengths are much larger than its size. This infrared divergence is not due to the pathological character of a massless scalar field in 
$(1+1)$-dimensional spacetime \cite{Coleman73};  it also appears in higher dimensional spacetime \cite{Frolov99}
if the mirror is accelerated for an infinitely  long  time. 
\begin{figure}[ht!]
\includegraphics[width=8.0cm]{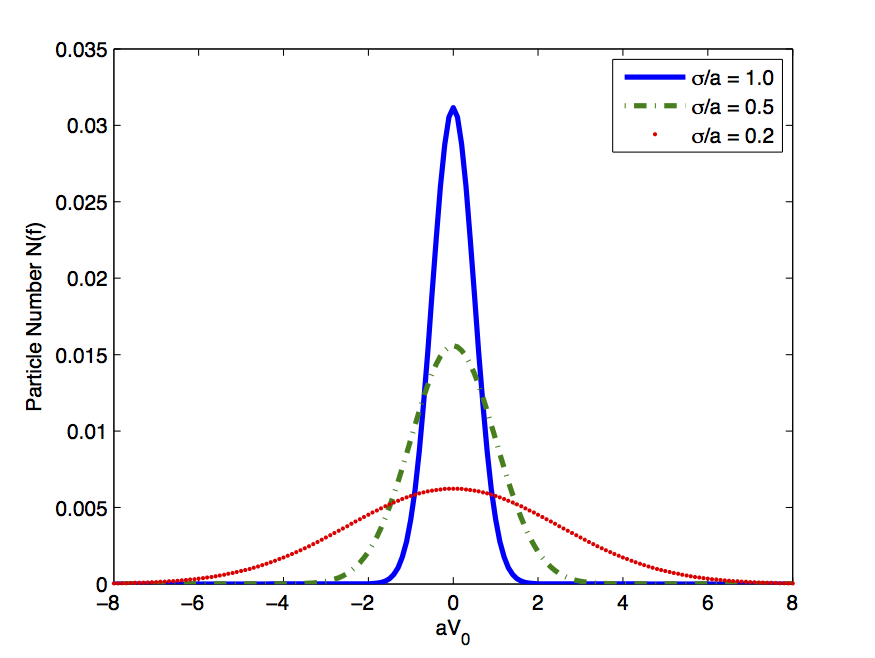}
\caption{\footnotesize (color online). Particle number versus central position of the Gaussian wave packet: $k_0/a = 20, a g = 10$. For 
larger bandwidth (narrower wave packet in time domain), the particle number distribution is narrower, showing that particles 
are localized around the past event horizon. } 
\label{Fig:pn}
\end{figure}

\begin{figure}[ht!]
\includegraphics[width=8.0cm]{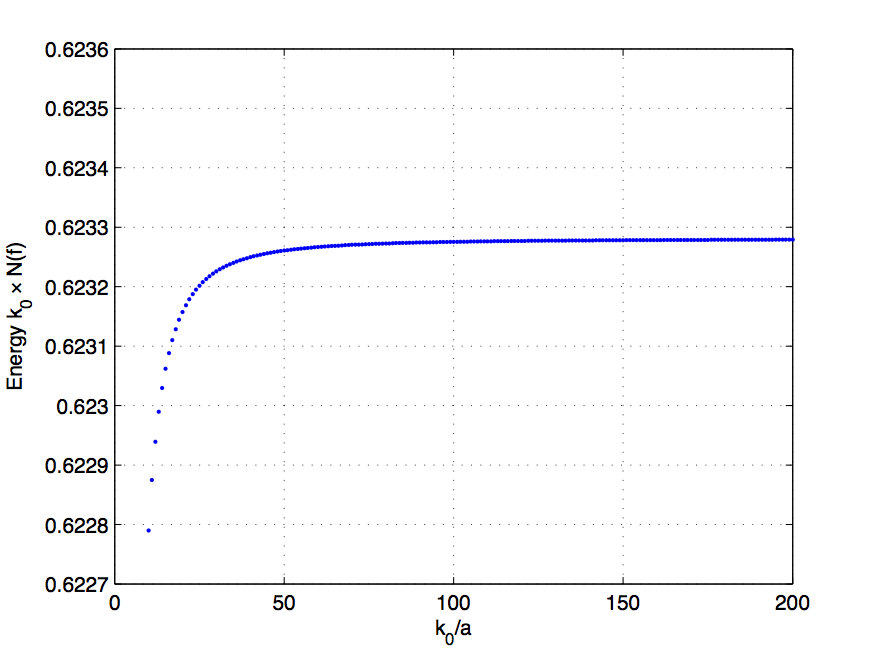}
\caption{\footnotesize (color online). Energy of the wave packets versus the central frequency: $\sigma/a = 1.0, a V_0 =0, a g = 10$. The energy is almost
constant in the high central frequency limit. }  
\label{Fig:energy}
\end{figure}

If we assume that the reflectivity $R_{\omega}$ of the mirror is a power law of $\omega$ as $\omega \rightarrow 0$ ($R_{\omega} \sim \omega^{\gamma}$)
then in order to obtain finite particle number we must have $\gamma > 1$.
As a concrete example, we choose 
\begin{equation}
R_{\omega} = \sin^2 \theta_{\omega} = \frac{g^2\omega^2}{1+g^2\omega^2},
\end{equation}
where $g$ is a parameter characterizing the low-frequency cutoff. Fig. \ref{Fig:pn} shows the particle number $N(f)$ versus 
the central position of the Gaussian wave packet. We can see that the particle-number distribution is symmetric with respect to $V_0 = 0$. 
In addition, for larger bandwidth (narrower wave packet in time domain), the distribution is more localized around $V_0 = 0$. These 
two facts indicate that the particle flux radiated by the uniformly accelerated mirror is well localized around the past horizon $V_0 = 0$. 
Since the mirror starts to accelerate in the distant past, that means the mirror only radiates particles  when it starts accelerating. 
It radiates no particles when it is uniformly accelerating. 

 Although in Eq. (\ref{ParticleNumber}) the integrand explicitly depends on the central frequency $k_0$ of the Gaussian wave packet, in the large $k_0$ limit
 the integration 
turns out to be almost independent of $k_0$.  That means the particle number $N(f) \sim \frac{1}{k_0}$ in the large
central frequency limit (see Appendix \ref{appendixC}), yielding  the relationship $E(f) \approx k_0 N(f) \sim \mathcal{O}(1)$, 
for the energy of the wave packet, 
as shown in 
Fig. \ref{Fig:energy}. Adding up the energy of 
all wave packets yields a divergent result. This ultraviolet divergence arises 
as a consequence of the physically unrealistic assumption that the mirror is accelerated eternally, so that  it appears to any  inertial observers when they cross the past horizon. This ultraviolet divergence can be removed by smoothly switching on the
mirror \cite{Obadia01}, or by considering an accelerated mirror whose acceleration was slowly increased from zero.
For a switch-on timescale of $\Delta T$, the particle number is suppressed for wave packets with central frequency $k_0 > \frac{1}{\Delta T}$ while
it remains the same for wave packets with central frequency $k_0 < \frac{1}{\Delta T}$. Therefore Eq. (\ref{ParticleNumber})
is not applicable to wave packets with very high central frequency  because it does not take into account physical initial conditions.

\section{Squeezing from accelerated mirrors}\label{squeezing}


A well known mechanism for generating particles from the vacuum is the two-mode squeezing process. Examples include non-degenerate 
parametric down conversion \cite{BachorRalph} and the Unruh effect \cite{Unruh76}. 
The two output modes are entangled with each other so that the 
composite state is a pure state. Another important mechanism is the single-mode squeezing process, for example degenerate parametric
down conversion \cite{BachorRalph}. It is possible that a particle generation process is the combination of the two, which we now show is the case for the uniformly 
accelerated mirror.  Using the quantum circuit model for 
the uniformly accelerated mirror, it is very easy to show that the wavepacket mode is squeezed at some quadrature phase depending on the central frequency and central
position of the wave packet. 

The correlations between various output Unruh modes are summarized in Appendix \ref{appendixB}. 
If we consider left-moving and narrow bandwidth Gaussian wave packet modes, using Eqs. (\ref{MinkowskiUnruh}), (\ref{GaussianMode}), (\ref{A}), (\ref{B}) and  (\ref{B1}), 
we have
\begin{widetext}
\begin{eqnarray}\label{CroCorrelation}
&&\langle 0_M |\hat{a}(f) \hat{a}(f) |0_M \rangle 
 = \int dk \int dk^{\prime} f(k) f(k^{\prime})  \int d\omega \int d\omega^{\prime} 
 \big[A_{k\omega} B_{k^{\prime} \omega^{\prime}}\langle 0_M |\hat{c}^{\prime}_{\omega} \hat{d}^{\prime}_{\omega^{\prime}} |0_M \rangle
+  B_{k\omega} A_{k^{\prime} \omega^{\prime}}\langle 0_M |\hat{d}^{\prime}_{\omega} \hat{c}^{\prime}_{\omega^{\prime}} |0_M \rangle\big] \nonumber \\
\nonumber \\
&=& - \sqrt{\frac{8}{\pi}}\frac{\sigma}{k_0} e^{-2ik_0V_0} \int_0^{\infty} d \Omega 
\exp\bigg[-\frac{\sigma^2(\Omega-k_0 V_0)^2}{k_0^2}\bigg] \exp\bigg[-\frac{\sigma^2(\Omega+k_0 V_0)^2}{k_0^2}\bigg]
(1-\cos \theta_{\Omega})e^{\pi \Omega}\frac{e^{2\pi \Omega}+1}{(e^{2\pi \Omega}-1)^2}. \nonumber \\
\end{eqnarray}
\end{widetext}
The quadrature observable of the localized wave packet mode $\hat{a}(f)$ is defined as 
\begin{equation}
\hat{X}(\phi) \equiv \hat{a}(f)e^{-i\phi} + \hat{a}^{\dagger}(f)e^{i \phi},
\end{equation}
where $\phi$ is the quadrature phase. 
From Eqs. (\ref{ParticleNumber}) and (\ref{CroCorrelation}), we find that for a narrow bandwidth Gaussian wave packet the variance is 
\begin{widetext}
\begin{eqnarray}\label{VarianceGaussian}
&&\big(\Delta X({\phi})\big)^2 = \langle 0_M | \hat{X}^2({\phi}) |0_M \rangle - \langle 0_M | \hat{X}({\phi}) |0_M \rangle^2 
 = 1+ 2 \langle 0_M | \hat{a}^{\dagger}(f)\hat{a}(f) |0_M \rangle + 2~ \text{Re} \bigg[\langle 0_M | \hat{a}(f)\hat{a}(f) |0_M \rangle e^{-2i\phi} \bigg] \nonumber \\
&=& 1+ 4\sqrt{\frac{2}{\pi}}\frac{\sigma}{k_0} \int_0^{\infty} d \Omega 
\bigg\{\exp\bigg[-\frac{2\sigma^2(\Omega-k_0 V_0)^2}{k_0^2}\bigg]+\exp\bigg[-\frac{2\sigma^2(\Omega+k_0 V_0)^2}{k_0^2}\bigg] \bigg\}
(1-\cos \theta_{\Omega})\frac{e^{2\pi \Omega}}{(e^{2\pi \Omega}-1)^2}\nonumber \\
&&- 4\sqrt{\frac{2}{\pi}}\frac{\sigma}{k_0} \cos(2\phi+2k_0V_0)\int_0^{\infty} d \Omega 
\exp\bigg[-\frac{\sigma^2(\Omega-k_0 V_0)^2}{k_0^2}\bigg] \exp\bigg[-\frac{\sigma^2(\Omega+k_0 V_0)^2}{k_0^2}\bigg]
(1-\cos \theta_{\Omega})e^{\pi \Omega}\frac{e^{2\pi \Omega}+1}{(e^{2\pi \Omega}-1)^2}, \nonumber \\ 
\end{eqnarray}
\end{widetext}
where we have used the fact that in the Minkowski vacuum state, $\langle 0_M | \hat{X}({\phi}) |0_M \rangle = 0$.
The variance of the wave packet mode could be smaller than one if the third term of Eq. (\ref{VarianceGaussian}) is 
larger than the second term. 
In order to show that single-mode squeezing is possible, we consider a Gaussian wave packet
centered at $V_0 = 0$. Eq. (\ref{VarianceGaussian}) considerably simplifies, yielding 
\begin{widetext}
\begin{eqnarray}
\big(\Delta X^{min}\big)^2 = 1 - 4\sqrt{\frac{2}{\pi}}\frac{\sigma}{k_0} \int_0^{\infty} d \Omega 
\exp \bigg(-\frac{2\sigma^2\Omega^2}{k_0^2}\bigg) 
\times (1-\cos \theta_{\Omega})\frac{e^{\pi \Omega}}{(e^{\pi \Omega}+1)^2} 
<1
\end{eqnarray} 
\end{widetext}
for the minimum of 
$\big(\Delta X({\phi})\big)^2$, which is at $\phi = 0$.

\begin{figure}[ht!]
\includegraphics[width=8.0cm]{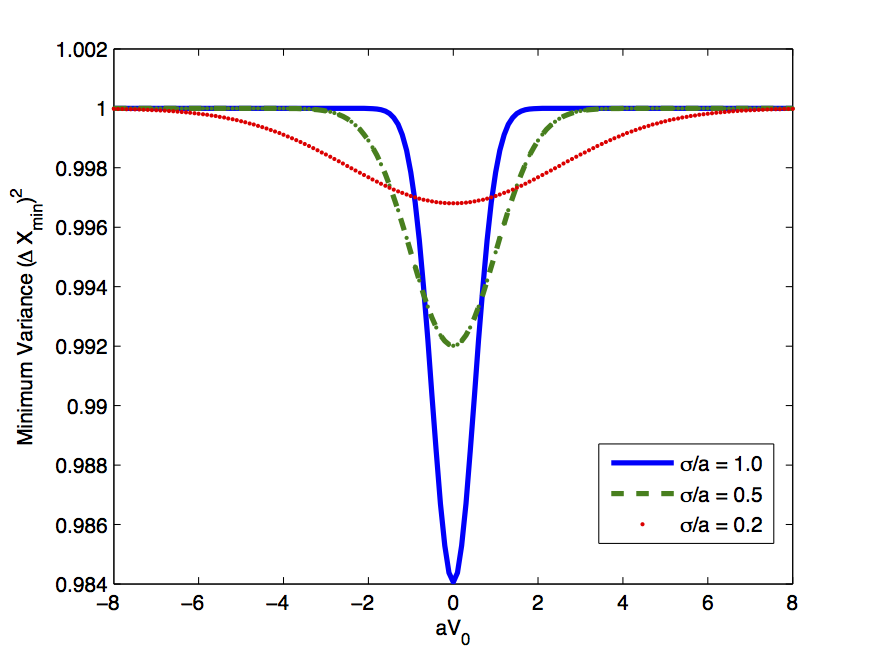}
\caption{\footnotesize (color online). Minimum variance versus central position of the Gaussian wave packet: $k_0/a = 20, a g = 10$. 
Maximum squeezing is achieved when the wave packet centers on the past horizon $V_0 = 0$. 
The squeezing is stronger for larger bandwidth wave packets.
} 
\label{Fig:variance}
\end{figure}

The variance of the quadrature beats the quantum shot noise, showing that the Gaussian wave packet mode is squeezed. 
When the center of the Gaussian wave packet is away from the past horizon $V_0 = 0$, the mode is squeezed at a different 
quadrature phase angle. According to Eq. (\ref{VarianceGaussian}), the minimum of the variance is reached when 
$\phi_s + k_0 V_0 =0$ is satisfied, that is 
\begin{equation}\label{phase}
\phi_s = - k_0 V_0. 
\end{equation}
The squeezing phase angle $\phi_s$ depends on both the central frequency and central position of the Gaussian wave packet. 
Other than the rotation of the squeezing phase angle, the squeezing amplitude decreases when the center of the wave packet is 
away from the past horizon. Fig. \ref{Fig:variance} shows the minimum variance of various wave packet modes
 (different central position and bandwidth), where the 
condition (\ref{phase}) has been satisfied.

From Fig. \ref{Fig:variance} we  see that the squeezing is stronger for a larger bandwidth Gaussian wave 
packet, which implies that different single-frequency Minkowski modes are also correlated. This can be verified if we replace
$f(k)$ in Eq. (\ref{CroCorrelation}) by a Dirac delta function $\delta(k - k_0)$. 
For a very large bandwidth wave packet  mode (such as  a broad bandwidth tophat mode), we find that   the minimum variance approaches  but never exceeds 0.5.  
We also note that when $\cos(2\phi + 2k_0 V_0) = -1$, the variance is maximal and larger than unity.

According to the quantum circuit model, it is easy to understand the origin of the single-mode squeezing. In Fig. \ref{circuit:Fig}, after
passing through the mirror the left-moving Rindler mode $\hat{b}_{\omega}^{\prime R}$ in the $R$ wedge is in thermal state, 
as well as the left-moving Rindler mode $\hat{b}_{\omega}^{L}$ in the $L$ wedge. The entanglement between 
$\hat{b}_{\omega}^{\prime R}$ and  $\hat{b}_{\omega}^{L}$ depends on the transmission coefficient of the mirror. 
If the mirror is completely transparent, they are perfectly entangled; while if the mirror is perfect,
the entanglement is completely severed. The Rindler modes $\hat{b}_{\omega}^{\prime R}$ and  $\hat{b}_{\omega}^{L}$
further pass through a two-mode antisqueezer $S^{-1}_{\omega}$, ending up with two Unruh modes 
$\hat{c}^{\prime}_{ \omega}$ and  $\hat{d}^{\prime}_{ \omega}$, which are also entangled. 
The amount of entanglement between $\hat{c}^{\prime}_{ \omega}$ and  $\hat{d}^{\prime}_{ \omega}$ depends on the amount of entanglement between
$\hat{b}_{\omega}^{\prime R}$ and  $\hat{b}_{\omega}^{L}$. If $\hat{b}_{\omega}^{\prime R}$ and  $\hat{b}_{\omega}^{L}$
are perfectly entangled, there is no entanglement between $\hat{c}^{\prime}_{ \omega}$ and  $\hat{d}^{\prime}_{ \omega}$;
otherwise, $\hat{c}^{\prime}_{ \omega}$ and  $\hat{d}^{\prime}_{ \omega}$ are partially entangled. 
From Eq. (\ref{MinkowskiUnruh}), the Minkowski mode $\hat{a}_k$ is a linear combination of the Unruh modes 
$\hat{c}^{\prime}_{ \omega}$ and  $\hat{d}^{\prime}_{ \omega}$. It is a general result in quantum optics that a linear combination
of entangled modes would produce single-mode squeezing, e.g., a $50:50$ beamsplitter transforms a two-mode squeezed
state into single-mode squeezed sate in each output mode. Therefore, the Minkowski mode $\hat{a}_k$ is squeezed. 

It is clear that the single-mode squeezing is closely related to the correlations across the horizon. 
If the mirror is transparent ($\cos \theta_{\Omega}  = 1$),
the correlations across the horizon are preserved and there is no single-mode squeezing. When one uses
a partially transmitting mirror ($\cos \theta_{\Omega}  < 1$) to sever the correlations across the horizon, 
single-mode squeezing is inevitably produced according to Eq. (\ref{VarianceGaussian}).

\section{Squeezed Firewall ?}\label{firewall}

Recently   three assertions about black hole evaporation were shown to be mutually  inconsistent\cite{AMPS}: (i) Hawking 
radiation is a unitary process, (ii) low energy effective field theory is valid near the event horizon, and (iii) an infalling observer encounters nothing 
unusual at the horizon. One of the proposed solutions to this paradox is that the infalling observer burns up at the horizon. A black hole firewall forms 
at the horizon for an old black hole and the correlations across the horizon are severed. 

Recently this firewall state was modeled for a 
Rindler horizon in Minkowski spacetime by severing correlations across the horizon.  The response of an Unruh-DeWitt detector was seen to be finite \cite{Louko14}.  The correlations across the horizon are severed by requiring the 
Wightman function to be zero, disregarding the underlying dynamics. Furthermore, a low-frequency cutoff in the Wightman
function was introduced, implying that correlations between high-frequency modes are cut whilst correlations between low-frequency modes are preserved. 
This is a warm firewall.  

We propose that a uniformly accelerated mirror is a possible mechanism for generating a Rindler firewall. From the quantum circuit
model we can see that the accelerated mirror acts as a pair of scissors cutting the correlations across the past horizon. If the mirror is perfect,
the correlations across the horizon are completely severed and the particle flux along the horizon is divergent. This is a hot firewall, destroying 
everything that crosses it. However, if the mirror is not perfect but transparent for low-frequency modes, the high-frequency correlations are 
cut while low-frequency correlations are preserved, and the particle flux in a localized wave packet mode along the horizon is finite, similar to the  warm firewall 
proposed by Louko \cite{Louko14}. In Sec. \ref{squeezing}, we showed that the radiation field from the accelerated mirror is squeezed,
which implies that the Rindler firewall is squeezed. It seems that squeezing is a general property of a Rindler firewall because in order to 
form a firewall one has to cut the correlations across the horizon, which inevitably generates single-mode squeezing.

Is a black hole firewall squeezed? Black hole firewalls are introduced in order to preserve the unitarity of  black hole evolution \cite{AMPS, Braunstein13}. 
For an old black hole, the late time Hawking radiation should be correlated with early time Hawking radiation but not with the degrees of freedom
inside the event horizon. The correlations across the horizon are severed during the evaporation. According to the arguments for
the Rindler firewall, it is reasonable to conjecture that the black hole firewalls are also squeezed. In addition, if the single-mode squeezing is 
strong enough, black hole firewalls do not have to be entangled with other unknown systems. 

\section{Conclusions}\label{conclusion}

We have developed a quantum circuit formalism to describe unitary interactions between a uniformly accelerated object and the quantum fields. 
The key point is to work in the accelerated frame where the object is stationary and couples only to Rindler modes in one of the Rindler wedges. 
If the initial state of the quantum fields is given in the inertial frame and the response of inertial detectors is considered, 
we have to transform modes from the inertial 
frame to the accelerated frame, which turns out to be a two-mode squeezing operation if we consider Unruh modes instead of Minkowski
modes in the inertial frame. We thus can construct a quantum circuit using two-mode squeezers and devices depending on the 
interaction of the object with the Rindler modes.

As an example, we studied a uniformly accelerated mirror. 
In the accelerated frame, the mirror is stationary and is simply a beamsplitter with frequency dependent reflection coefficient. The input-output
relation of a beamsplitter is well known and is widely used in quantum optics \cite{BachorRalph}. 
The quantum circuit for the uniformly accelerated mirror is shown in Fig. \ref{circuit:Fig}. As an application,
we calculated the radiation flux from an eternally accelerating mirror in the Minkowksi vacuum. We found that the particles are localized
around the horizon and the particle number in a localized wave packet mode is divergent if no low frequency regularization is introduced. 

Our results are consistent with earlier results obtained using different methods \cite{Frolov99, Obadia01}. The infrared divergence occurs due to the ideal assumption that the mirror 
accelerates for an infinitely long time. We emphasize that the infrared divergence is not due to the particular pathological character of a massless scalar field in 
$(1+1)$-dimensional spacetime \cite{Coleman73} because it also appears in higher dimensional spacetime \cite{Frolov99}.
We regularize the radiation flux by introducing a low-frequency cutoff for the mirror, that is, the mirror is completely transparent
for low frequency field modes. Physically, this is equivalent to having the mirror interact with the field for a finite time.  
After regularizing the infrared divergence, the particle number of a localized wave packet mode is finite. However the energy of the wave packet mode does not decay as the central frequency increases, in turn implying that the total energy of the radiation flux is infinite. This ultraviolet divergence
arises because of the naive assumption that the mirror  is accelerated eternally so that it appears to inertial observers when they cross the past horizon. If the mirror slowly increased its 
acceleration or was switched on smoothly, the number of high frequency particles would be suppressed, removing this ultraviolet divergence. 
Using  perturbation theory it is straightforward to show
that the energy flux is finite if the mirror is smoothly turned on and off \cite{Obadia01}. 

A further application of our circuit model would be in the study a uniformly accelerated cavity. Previous work on this topic \cite{Alsing03,Downes11,Bruschi12} studied
how the quantum states stored inside a perfect cavity are affected by   uniform acceleration. 
While  Unruh-Davies radiation \cite{Unruh76, Davies75} cannot affect the field modes inside a perfect cavity,  it can affect field modes inside
an imperfect one. Because the circuit model is designed to study an imperfect uniformly accelerated mirror, we believe that by generalizing 
the model from one mirror to two mirrors, one can study the interaction between Unruh-Davies radiation and the field modes inside an imperfect cavity. 

One limitation of our circuit model is that it is only suitable for studying hyperbolic trajectories in Minkowski spacetime;  more general trajectories are not straightforwardly incorporated. 
One might expect this to severely limit the utility of  the circuit model because physically it is not possible to accelerate a mirror for an infinitely long time. However our use of the transparency term shows that we  
can turn on and off the mirror so that it is transparent in the distant past and distant future. 
This could be used to model a mirror that  initially undergoes inertial motion, accelerates for a finite period of time, and then returns to inertial motion.
We will leave this topic for future work.


We find that the radiation flux from the uniformly accelerated mirror is squeezed.  
To the best of our knowledge, the contribution of single-mode squeezing to the generation of particles by a moving mirror  has not been discussed previously. 
The squeezing angle depends on the central frequency and 
position of the localized detector mode function. Maximum squeezing occurs when the detector mode function centers on the 
horizon.  It is clear from the circuit model that the squeezing 
is related to the correlations across the horizon. When the mirror is completely transparent, the correlations across the horizon are preserved
and there is no squeezing. When the mirror completely reflects a Rindler mode with a particular frequency, it destroys the correlation across
the horizon and generates some squeezing in the Minkowski mode. It therefore provides a mechanism for transferring the correlations
across the horizon to the squeezing of the radiation flux on the horizon. 

Recently, Louko \cite{Louko14} proposed a Rindler firewall state by severing the 
correlations across the horizon by hand and claimed that the response of a particle detector is finite. It was subsequently shown that entanglement 
survives this Rindler firewall \cite{Martinez15}. Our calculation suggests that one way of generating a Rindler firewall is to uniformly accelerate
a mirror.  
We conjecture that if the firewall is formed in an old black hole, the radiation flux at the horizon
could be squeezed as the price of severing the entanglement across the event horizon. In addition, the black hole firewall may not need to be highly entangled with 
other systems \cite{Susskind14} because the squeezing may be enough to account for the particle flux on the horizon.

\section*{ACKNOWLEGEMENTS}
We would like to thank Antony Lee, Shih-Yuin Lin and Yiqiu Ma for useful discussions. 
This research was supported in part by Australian Research Council Centre of Excellence of Quantum 
Computation and Communication Technology (Project No. CE110001027),
and in part by the Natural Sciences and Engineering Research Council of Canada.

\appendix
\vspace{0.5cm}
\section{Input-output relations\label{appendixA}}

We summarize the input-output relations of the quantum circuit Fig. \ref{circuit:Fig} with the object a beamsplitter.
The action of the beamsplitter is represented by Eq. (\ref{mirrortransformation}). Substituting it into Eq. (\ref{input-output}), we have
\begin{widetext}
\begin{eqnarray}\label{UnruhLeftC}
\hat{\bf c}^{\prime}_{1\omega} &=& \hat{\bf c}_{1\omega}[\text{cosh}^2(r_{\omega})\text{cos} \,\theta_{\omega}-\text{sinh}^2(r_{\omega})] 
- \sigma_x  \hat{\bf d}_{1\omega} \text{cosh}(r_{\omega})\text{sinh}(r_{\omega})(1-\text{cos} \,\theta_{\omega})
+Z  \hat{\bf c}_{2\omega} \text{cosh}^2(r_{\omega})\text{sin} \,\theta_{\omega} \nonumber \\
&&+Z \sigma_x \hat{\bf d}_{2\omega}  \text{cosh}(r_{\omega})\text{sinh}(r_{\omega}) \text{sin} \,\theta_{\omega}. \nonumber \\
&=& [\text{cosh}^2(r_{\omega})\text{cos} \,\theta_{\omega}-\text{sinh}^2(r_{\omega})] {\hat{c}_{1\omega} \choose \hat{c}_{1\omega}^{\dag}}
-  \text{cosh}(r_{\omega})\text{sinh}(r_{\omega})(1-\text{cos} \,\theta_{\omega}){\hat{d}_{1\omega}^{\dag} \choose \hat{d}_{1\omega}} 
+ \text{cosh}^2(r_{\omega})\text{sin} \,\theta_{\omega} {-i e^{i \phi_{\omega}}\hat{c}_{2\omega} \choose i e^{- i \phi_{\omega}}\hat{c}_{2\omega}^{\dag}} \nonumber \\
&&+ \text{cosh}(r_{\omega})\text{sinh}(r_{\omega}) \text{sin} \,\theta_{\omega}  {-i e^{i \phi_{\omega}}\hat{d}_{2\omega}^{\dag} \choose i e^{- i \phi_{\omega}}\hat{d}_{2\omega}}, 
\end{eqnarray}
\
\begin{eqnarray}\label{UnruhLeftD}
\hat{\bf d}^{\prime}_{1\omega} &=& \sigma_x  \hat{\bf c}_{1\omega} \text{cosh}(r_{\omega})\text{sinh}(r_{\omega})(1-\text{cos}\,\theta_{\omega})
+ \hat{\bf d}_{1\omega}[\text{cosh}^2(r_{\omega})-\text{sinh}^2(r_{\omega})\text{cos}\,\theta_{\omega}] 
- \sigma_x Z \hat{\bf c}_{2\omega}  \text{cosh}(r_{\omega})\text{sinh}(r_{\omega}) \text{sin}\,\theta_{\omega} \nonumber \\
&& - \sigma_x Z  \sigma_x \hat{\bf d}_{2\omega} \text{sinh}^2(r_{\omega})\text{sin}\,\theta_{\omega}  \nonumber \\
&=& \text{cosh}(r_{\omega})\text{sinh}(r_{\omega})(1-\text{cos}\,\theta_{\omega}){\hat{c}_{1\omega}^{\dag} \choose \hat{c}_{1\omega}}
+[\text{cosh}^2(r_{\omega})-\text{sinh}^2(r_{\omega})\text{cos}\,\theta_{\omega}]{\hat{d}_{1\omega} \choose \hat{d}_{1\omega}^{\dag}}  \nonumber \\
&&- \text{cosh}(r_{\omega})\text{sinh}(r_{\omega}) \text{sin}\,\theta_{\omega}{i e^{- i \phi_{\omega}}\hat{c}_{2\omega}^{\dag} \choose -i e^{i \phi_{\omega}}\hat{c}_{2\omega}} 
- \text{sinh}^2(r_{\omega})\text{sin}\,\theta_{\omega}{ i e^{- i \phi_{\omega}}\hat{d}_{2\omega} \choose -i e^{i \phi_{\omega}}\hat{d}_{2\omega}^{\dag} },  
\end{eqnarray}
\
\begin{eqnarray}\label{UnruhRightC}
\hat{\bf c}^{\prime}_{2\omega} &=& -Z^*  \hat{\bf c}_{1\omega} \text{cosh}^2(r_{\omega})\text{sin}\,\theta_{\omega} 
-Z^* \sigma_x \hat{\bf d}_{1\omega}  \text{cosh}(r_{\omega})\text{sinh}(r_{\omega}) \text{sin}\,\theta_{\omega}
+ \hat{\bf c}_{2\omega}[\text{cosh}^2(r_{\omega})\text{cos}\,\theta_{\omega}-\text{sinh}^2(r_{\omega})] \nonumber \\
&&-\sigma_x  \hat{\bf d}_{2\omega} \text{cosh}(r_{\omega})\text{sinh}(r_{\omega})(1-\text{cos}\,\theta_{\omega}) \nonumber \\
&=&\text{cosh}^2(r_{\omega})\text{sin}\,\theta_{\omega} {-ie^{-i \phi_{\omega}}\hat{c}_{1\omega} \choose ie^{i \phi_{\omega}}\hat{c}_{1\omega}^{\dag}}  
+ \text{cosh}(r_{\omega})\text{sinh}(r_{\omega}) \text{sin}\,\theta_{\omega} {-i e^{-i \phi_{\omega}}\hat{d}_{1\omega}^{\dag} \choose i e^{i \phi_{\omega}}\hat{d}_{1\omega}} 
+[\text{cosh}^2(r_{\omega})\text{cos}\,\theta_{\omega}-\text{sinh}^2(r_{\omega})] {\hat{c}_{2\omega} \choose \hat{c}_{2\omega}^{\dag}}\nonumber \\ 
&&-  \text{cosh}(r_{\omega})\text{sinh}(r_{\omega})(1-\text{cos}\,\theta_{\omega}){\hat{d}_{2\omega}^{\dag} \choose \hat{d}_{2\omega}},  
\end{eqnarray}
\
\begin{eqnarray}\label{UnruhRightD}
\hat{\bf d}^{\prime}_{2\omega} &=& -\sigma_x Z^* \hat{\bf c}_{1\omega}  \text{cosh}(r_{\omega})\text{sinh}(r_{\omega}) \text{sin}\,\theta_{\omega}
 + \sigma_x Z^*  \sigma_x \hat{\bf d}_{1\omega} \text{sinh}^2(r_{\omega})\text{sin}\,\theta_{\omega}
 + \sigma_x  \hat{\bf c}_{2\omega} \text{cosh}(r_{\omega})\text{sinh}(r_{\omega})(1-\text{cos}\,\theta_{\omega})  \nonumber \\
 &&+ \hat{\bf d}_{2\omega}[\text{cosh}^2(r_{\omega})-\text{sinh}^2(r_{\omega})\text{cos}\,\theta_{\omega}] \nonumber \\
 &=&-\text{cosh}(r_{\omega})\text{sinh}(r_{\omega}) \text{sin}\,\theta_{\omega}{i e^{ i \phi_{\omega}}\hat{c}_{1\omega}^{\dag} \choose -i e^{-i \phi_{\omega}}\hat{c}_{1\omega}} 
- \text{sinh}^2(r_{\omega})\text{sin}\,\theta_{\omega}{ i e^{ i \phi_{\omega}}\hat{d}_{1\omega} \choose -i e^{-i \phi_{\omega}}\hat{d}_{1\omega}^{\dag} } \nonumber \\
&&+\text{cosh}(r_{\omega})\text{sinh}(r_{\omega})(1-\text{cos}\,\theta_{\omega}) {\hat{c}_{2\omega}^{\dag} \choose \hat{c}_{2\omega}}
+[\text{cosh}^2(r_{\omega})-\text{sinh}^2(r_{\omega})\text{cos}\,\theta_{\omega}]{\hat{d}_{2\omega} \choose \hat{d}_{2\omega}^{\dag}} \label{UnruhRightD}.
\end{eqnarray}

\end{widetext}

\section{Correlations between output Unruh modes\label{appendixB}}

Using Eqs. (\ref{UnruhLeftC})-(\ref{UnruhRightD}), it is straightforward to calculate the correlations between various output
Unruh modes in the Minkowski vacuum state. 
\begin{widetext}

\begin{eqnarray}\label{B1}
\langle 0_M | \hat{c}'_{m \omega}  \hat{d}'_{m \omega'} |0_M \rangle &=& \langle 0_M | \hat{d}'_{m \omega}  \hat{c}'_{m \omega'} |0_M \rangle
= \langle 0_M | \hat{c}^{\prime \dag}_{m \omega}  \hat{d}^{\prime \dag}_{m \omega'} |0_M \rangle 
= \langle 0_M | \hat{d}^{\prime \dag}_{m \omega}  \hat{c}^{\prime \dag}_{m \omega'} |0_M \rangle \nonumber \\
\nonumber \\
&=& - (1 - \cos \theta_{\omega}) \cosh(r_{\omega}) \sinh(r_{\omega}) \bigg [\sinh^2(r_{\omega}) + \cosh^2(r_{\omega})  \bigg] \delta(\omega - \omega'),
\end{eqnarray}

\begin{eqnarray}\label{B2}
\langle 0_M | \hat{c}'_{1 \omega}  \hat{d}'_{2 \omega'} |0_M \rangle &=& \langle 0_M | \hat{d}'_{2 \omega}  \hat{c}'_{1 \omega'} |0_M \rangle
= \langle 0_M | \hat{c}^{\prime \dag}_{1 \omega}  \hat{d}^{\prime \dag}_{2 \omega'} |0_M \rangle ^*
= \langle 0_M | \hat{d}^{\prime \dag}_{2 \omega}  \hat{c}^{\prime \dag}_{1 \omega'} |0_M \rangle^* \nonumber \\
\nonumber \\
&=& i e^{i \varphi_{\omega}}\sin \theta_{\omega} \cosh(r_{\omega}) \sinh(r_{\omega}) \delta(\omega - \omega'),
\end{eqnarray}

\begin{eqnarray}\label{B3}
\langle 0_M | \hat{c}'_{2 \omega}  \hat{d}'_{1 \omega'} |0_M \rangle &=& \langle 0_M | \hat{d}'_{1 \omega}  \hat{c}'_{2 \omega'} |0_M \rangle
= \langle 0_M | \hat{c}^{\prime \dag}_{2 \omega}  \hat{d}^{\prime \dag}_{1 \omega'} |0_M \rangle ^*
= \langle 0_M | \hat{d}^{\prime \dag}_{1 \omega}  \hat{c}^{\prime \dag}_{2 \omega'} |0_M \rangle^* \nonumber \\
\nonumber \\
&=& - i e^{- i \varphi_{\omega}}\sin \theta_{\omega} \cosh(r_{\omega}) \sinh(r_{\omega}) \delta(\omega - \omega'),
\end{eqnarray}

\end{widetext}
with others zero and here $m = 1,2$. 

\section{High central frequency limit}\label{appendixC}

We derive an analytically approximate expression for the particle number $N(f)$ in the high central frequency limit. 
From Eq. (\ref{ParticleNumber}), one expects that the term in the braces has two peaks at $k_0 V_0$ and $-k_0 V_0$. If $k_0$ is large
then the peaks are far away from the origin. However, the factor $\frac{e^{2\pi \Omega}}{(e^{2\pi \Omega}-1)^2}$ exponentially decays
for large $\Omega$ so that it strongly suppresses one of the Gaussian peaks. Therefore, the main contribution to the integration is from the low 
frequency. We Taylor expand the term in the braces to second order,
\begin{widetext}
\begin{eqnarray}
\exp\bigg[-\frac{2\sigma^2(\Omega-k_0 V_0)^2}{k_0^2}\bigg]+\exp\bigg[-\frac{2\sigma^2(\Omega+k_0 V_0)^2}{k_0^2}\bigg] 
\approx 2 e^{-2 \sigma^2 V_0^2} + \frac{4 \sigma^2 \Omega^2}{k_0^2} (4 \sigma^2 V_0^2 -1) e^{-2 \sigma^2 V_0^2}. 
\end{eqnarray}
\end{widetext}
In order to get an analytic expression, 
we introduce sharp low frequency cutoff, $R_{\omega} = 1$ for $\Omega \ge \epsilon$ and zero for $0 < \Omega < \epsilon$. Therefore we have
$1-\cos \theta_{\Omega} = 1$ for $\Omega \ge \epsilon$ and zero for $0 < \Omega < \epsilon$. The particle number $N(f)$ can be approximated as
\begin{widetext}

\begin{eqnarray}\label{ParticleNumber:2}
N(f) &\approx& 4 \sqrt{\frac{2}{\pi}}\frac{\sigma}{k_0} e^{-2 \sigma^2 V_0^2} 
\bigg[ \int_{\epsilon}^{\infty} d \Omega \frac{e^{2\pi \Omega}}{(e^{2\pi \Omega}-1)^2} +
\frac{2 \sigma^2}{k_0^2} (4 \sigma^2 V_0^2 -1) \int_{\epsilon}^{\infty} d \Omega \frac{\Omega^2 e^{2\pi \Omega}}{(e^{2\pi \Omega}-1)^2} \bigg] \nonumber \\
\nonumber \\
&\approx&  \bigg(\frac{2}{\pi} \bigg)^{3/2} \bigg(\frac{\sigma}{k_0} \bigg) e^{-2 \sigma^2 V_0^2} 
\bigg[\frac{1}{e^{2 \pi \epsilon} - 1} +
\frac{2 \sigma^2}{k_0^2} (4 \sigma^2 V_0^2 -1)\bigg(\frac{1}{12} - \frac{\epsilon^2}{2 \pi} \bigg) \bigg]. 
\end{eqnarray}

\end{widetext}
Comparison with direct numerical calculation shows that Eq. (\ref{ParticleNumber:2}) is a very good approximation when $\epsilon$ is small. 
We can see that the particle number is dependent on the low frequency cutoff $\epsilon$. The first term of Eq. (\ref{ParticleNumber:2}) is proportional
to $\frac{1}{e^{2 \pi \epsilon} - 1}$ which is divergent when $\epsilon \rightarrow 0$. Furthermore, in the high central frequency limit $k_0 \rightarrow \infty$,
the leading order of $N(f)$ is proportional to $\frac{1}{k_0}$.

\end{document}